\documentclass[preprint,12pt,authoryear]{elsarticle}
\newcommand{\apj}{ApJ}
\newcommand{\aj}{AJ}
\newcommand{\aap}{A\&A}
\newcommand{\apjs}{ApJS}
\newcommand{\mnras}{MNRAS}
\newcommand{\araa}{ARA\&A}

\usepackage{graphicx}
\usepackage{epsfig}

\usepackage{amssymb}

\usepackage[colorlinks=true, linkcolor=blue, citecolor=red, urlcolor=blue]{hyperref}
\usepackage{natbib}
\bibpunct{(}{)}{;}{a}{}{,}

\journal{New Astronomy Reviews}

\begin{document}

\begin{frontmatter}



\title{Studying the spectral properties of Active Galactic Nuclei in the JWST era}


\author{Th. Nakos, M. Baes}
\address{Sterrenkundig Observatorium, University of Ghent, B-9000, Belgium}
\author{A. Alonso-Herrero, A. Labiano}
\address{Departmento de Astrof\'isica Molecular e Infrarroja, IEM, CSIC, Madrid, Spain}

\begin{abstract}

The James Webb Space Telescope (JWST), due to launch in 2014, shall provide
an unprecedented wealth of information in the near and mid-infrared wavelengths, 
thanks to its high-sensitivity instruments and its 6.5 m primary mirror, 
the largest ever launched into space. NIRSpec and MIRI, the two spectrographs onboard JWST, 
will play a key role in the study of the spectral features of Active Galactic Nuclei in the $0.6-28$ micron wavelength range. This talk aims at presenting an overview of the possibilities provided by these two instruments, in order to prepare the astronomical community for the JWST era.

\end{abstract}


\end{frontmatter}



\section{Introduction}
\label{intro}

The James Webb Space Telescope is a project led by NASA, with major contributions from the European and Canadian Space Agency (ESA, CSA, respectively). Known initially as the Next Generation Space Telescope (NGST), it was renamed in 2002 as the James Webb Space Telescope, after the former NASA administrator. Meant to replace the ageing Hubble Space Telescope (HST), JWST is due to launch in 2014 on an Ariane 5 rocket, from the French Guiana. Its destination will be the semi-stable L2 Lagrange point, some $1.5 \times 10^6$\,km from Earth. The foreseen life of the mission is five years, although there is a possibility of extension up to ten years. 

JWST's primary mirror consists of 18 hexagonal, beryllium-made segments, resulting in a collecting area of 25\,m$^2$. As the size of the mirror (6.5\,m diameter), and that of other spacecraft elements, are too large to be accommodated in the rocket shrouds, JWST will be constructed using deployable structures. Thanks to its large aperture and high-sensitivity instruments, JWST is expected to extend the science done with HST not only in terms of performance (higher sensitivity, better resolution) but also in terms of wavelength range. JWST shall be diffraction limited at $2\,\mu$m and its instruments will be performing in the $0.6-29~\mu$m range, limited at the short end by the gold coatings of the primary mirror and at the long end by the mid-infrared (hereafter MIR) detector quantum efficiency. The JWST integrating science instrument module (ISIM) consists of four instruments: a near-infrared camera \cite[NIRCam,~][]{horner04}, a near-infrared spectrograph \cite[NIRSpec,~][]{zam04}, a mid-infrared instrument \cite[MIRI,~][]{wright04} and a near-infrared tunable filter imager \cite[TFI,~][]{rowlands04a}. A detailed description of the space observatory and its science cornerstones can be found in~\cite{gardner06}.

\section{The near-infrared spectrograph (NIRSpec)}
NIRSpec is a near-infrared spectrograph, operating in the $0.6-5\,\mu$m wavelength range, with NASA providing the two HgCdTe 2k$\times$2k detectors and the multi-slit system, and ESA the instrument. NIRSpec will be operating in three observing modes: 
\begin{itemize}
\item  Multi-object spectroscopy: thanks to a micro-shutter array (MSA, also provided by NASA), whose elements can either be closed or opened independently, it can simultaneously obtain up to 100 spectra, in a field of view (FoV) of $\approx 3.4 \times 3.4$ arcmin$^2$. 
\item Long-slit spectroscopy: six fixed slits (three slits of $3.5''\times 0.2''$, one $4''\times 0.4''$ and one $2''\times 0.1''$) will allow to reach a resolving power between $1000 < R < 2700$ in three spectral bands ($1.0 - 1.8\,\mu$m, $1.7 - 2.9\,\mu$m, $2.9 - 5.0\,\mu$m). Using a single prism, it will also be possible to cover the $0.6-5\,\mu$m wavelength range, but at a lower resolution ($R \approx 100$).
\item Integral Field Spectroscopy: NIRSpec includes an  integral field unit (IFU) with 30 slices, each 0.1$''$  wide, with a FoV of $3''\times3''$.

\end{itemize}
The resolution for the different spectroscopic modes is presented in Table~\ref{nirspec-tbl}.

\begin{table}[t]
\begin{center}
\begin{tabular}{cccc}
\hline\noalign{\smallskip}
R &
Prism/Grating &
$\lambda$\,($\mu$m) &
Mode \\[3pt] \hline
\noalign{\smallskip}
100  &  single prism  &  $0.6-5 $  & MOS, fixed slit\\
1000 &  3 gratings    &  $1 - 5 $  & MOS, fixed slit\\
3000 &  3 gratings    &  $1 - 5 $  & MOS, IFS\\
\noalign{\smallskip}\hline
\end{tabular}
\end{center}
\caption{\label{nirspec-tbl}NIRSpec observing modes: Multi-Object Spectroscopy (MOS), long-slit spectroscopy or Integral Field Spectroscopy (IFS) as a function of spectral resolution.}
\end{table}

\section{The mid-infrared instrument (MIRI)}
MIRI is a European-American collaboration and is unique, in many aspects. First of all, it will be the only instrument operating in the MIR wavelength range ($5 - 28\,\mu$m). Secondly, it is also the only JWST instrument that can perform imaging, spectroscopy and coronagraphy. MIRI consists of three identical $1024 \times 1024$ Si:As sensor chip assemblies (SCA), of 25~$\mu$m pixel size, that shall be operating at 7\,K. As the spacecraft will be passively cooled down to $\approx 35$\,K, the 7\,K temperature will be achieved by means of a mechanical cooler.  The MIRI SCAs and cryo-cooler are provided by Jet Propulsion Laboratory, while MIRI's optical system is provided by a consortium of 28 institutes from 10 European countries. 

MIRI's imager provides broad and narrow-band imaging, phase-mask coronagraphy, Lyot coronagraphy and low-resolution slit spectroscopy (LRS). The LRS covers the $5-10\,\mu$m range with a resolution of R$\,\approx 100$ at 7.5\,$\mu$m, using a slit width of 0.6$''$ and a length of 5.5$''$. The imager will have a pixel scale of 0.11$''$ and a FoV (excluding the region occupied by the coronagraphs and the LRS) of $79'' \times 113''$. 

\begin{figure}[h!]
\centering
\includegraphics[width=1.0\textwidth]{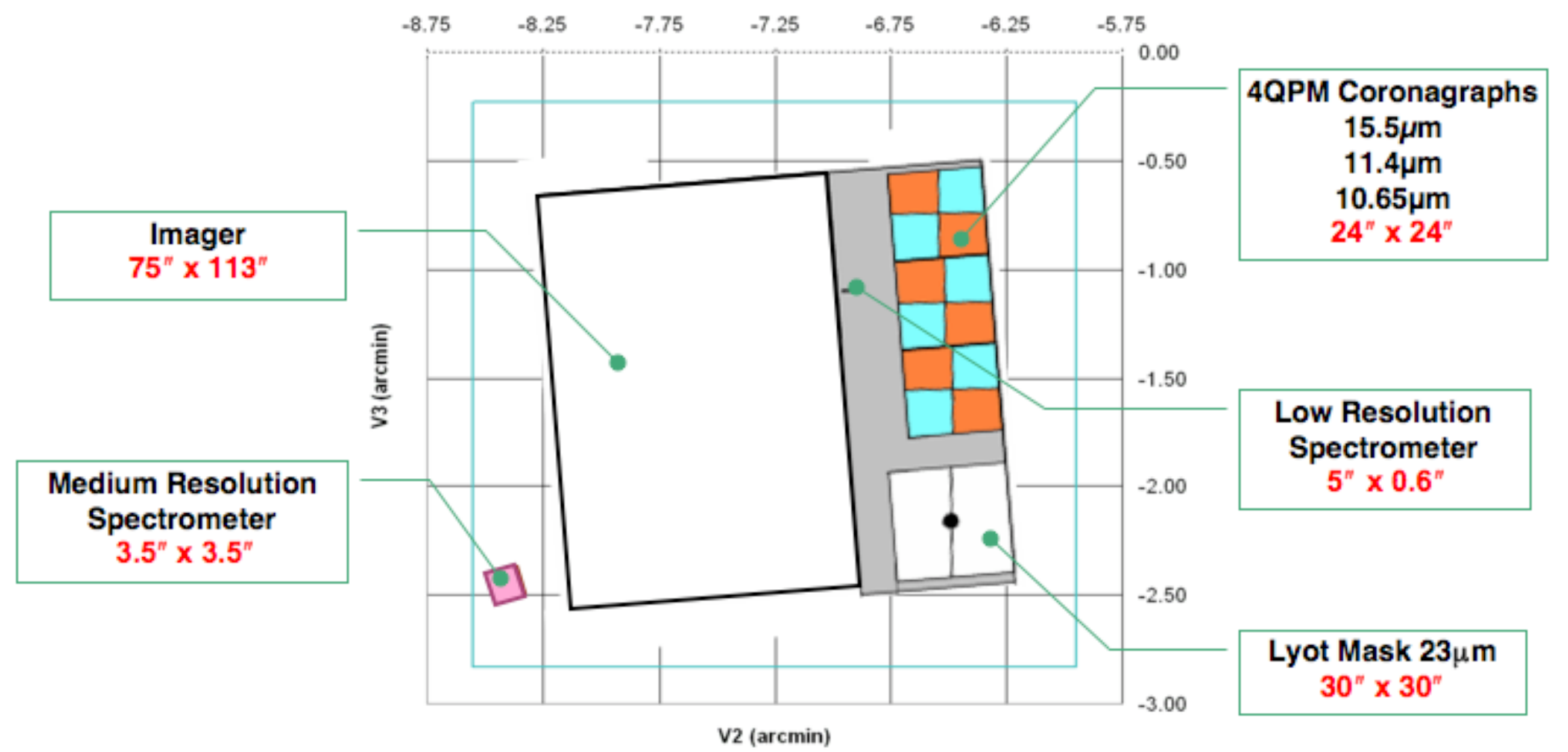}
\caption{The layout of the MIRI Imager focal plane and the Medium Resolution Spectrometer within the allocated region box) of the JWST focal plane. For the Imager the position of the three four-quadrant phase-mask (4QPM) coronagraphs, of the Lyot stop and of the low-resolution spectrograph, are also visible.}
\label{miri}
\end{figure}

MIRI's two IFUs will allow to derive spectral and spatial information of the targeted objects with a resolution of R\,$\approx 2000 - 3700$ over the $ 5\,\mu$m $< \lambda < 27\,\mu$m wavelength range. 
The IFUs provide four simultaneous and concentric FoVs, that increase in size with wavelength. 
One of the two IFUs will cover the short-wavelength range ($5-11.9\,\mu$m) while the second one will cover
the long-wavelength range ($11.9-28.3\,\mu$m). The main properties of the MIRI medium resolution spectrometer (MRS) are presented in Table~\ref{ifus-tbl}. Fig.~\ref{miri} shows how MIRI's imager and MRS will be occupying the allocated region of the JWST focal plane. 

\begin{table}[t]
{\footnotesize
\begin{tabular}{ccccccc}
\hline\noalign{\smallskip}
 $\lambda$   &  Slices  & \multicolumn{2}{c}{Spatial sample dimens.} & \multicolumn{2}{c}{FoV }             & {R} \\
 {$\mu$m}    &          &   Across ($''$) & Along  ($''$)         & Across   ($''$)      & Along ($''$)  &                 \\           
[3pt]
\noalign{\smallskip}
 5.0 $-$ 7.7  &   21    &    0.18       &  0.20                 & 3.7                 &  3.7         & 2400 $-$ 3700  \\
 7.7 $-$ 11.9 &   17    &    0.28       &  0.20                 & 4.5                 &  4.7         & 2400 $-$ 3600  \\
11.9 $-$ 18.3 &   16    &    0.39       &  0.25                 & 6.1                 &  6.2         & 2400 $-$ 3600  \\
18.3 $-$ 28.3 &   12    &    0.64       &  0.27                 & 7.9                 &  7.7         & 2000 $-$ 2400  \\
\noalign{\smallskip}\hline
\end{tabular}
\caption{Properties of the MIRI integral field units\label{ifus-tbl}. The region of the sky corresponding to a spatial sample is set by the slice width in the across slice direction, and by the pixel field of view in the along slice direction.}
}
\end{table}

\section{NIRSpec and MIRI sensitivity}
Thanks to its large mirror, the location from which the observatory will be operating (L2 point) and its high-sensitivity detectors, JWST will be at least 100 times more sensitive than Spitzer and will provide an order of magnitude improved spatial/spectral resolution. More specifically, for NIRSpec, the sensitivity requirements for the R $\approx 100$ mode is a flux continuum of 132 nJy (AB $>$ 28) at 3.0$\,\mu$m for a 10\,000 sec exposure and a signal-to-noise ratio of 10; for the R $\approx 1000$ mode the limiting line-flux will be $\approx 6\times10^{-19}$erg\,s$^{-1}$\,cm$^{-2}$ at 2$\,\mu$m~\citep{gardner06}. 
Regarding MIRI, the requirement sensitivity at 6.4, 9.2, 14.5 and 22.5$\,\mu$m for a 10\,000 sec exposure and a 10-$\sigma$ detection is 12.2, 9.9, 11.6 and 59.7$\times 10^{-21}$W/m$^2$, respectively~\citep{swin04}. 

\section{AGN-related science with NIRSpec \& MIRI}
\label{nirspec-miri-science}
Active Galactic Nuclei (AGN) play a major role in astrophysical research, for the following reasons:
\begin{itemize}
   \item Quasars, the most luminous members of the AGN family, are among the few cosmological light-houses that can probe the lower-end ($z \approx 6-7$) of the reionization epoch. 
   \item Non-quasar AGN, detected at lower redshifts ($z < 3$), are a key to understanding the cosmological evolution once the Universe became fully ionized: how galaxies formed, what triggers the black hole (BH) activity, what is the relation between starbursts and AGN, why observations contradict the predictions of the so-called AGN unification scheme. 
   \item The vicinity of the BH is one of the most complex physical laboratories, with electromagnetic radiation being emitted from the X-rays to the far-IR and even radio-waves. 
\end{itemize}
Spectroscopy is indispensable for shedding light on all issues mentioned above. In the following sub-sections we briefly present some of the aspects in which NIRSpec and MIRI will potentially have a major contribution.

\subsection{Probing the reionization epoch via high-z quasar spectra}
The detection of high-redshift quasars has proven to be a powerful tool for studying the low-end of the reionization epoch~\citep{fan06}, via the Gunn-Peterson trough~\citep{gp65}. Quasar-based studies imply that the Universe was opaque (i.e. contained a large fraction of neutral hydrogen) down to $z\approx 6$. On the other hand, CMB polarization studies suggest that the ``dark ages" ended at $z_{reion} = 10.9 \pm 1.4$, with the Universe changing instantaneously from neutral to a fully ionized state at this specific redshift~\citep{komatsu09}. 

A possible cause for this discrepancy might be due to (a) the assumptions made for the data analysis or (b) data-related uncertainties. More specifically, difficulties in obtaining robust redshifts estimates, the insufficient precision with which the latter ones are known, and uncertainties on the calculation of the depression of the quasar continuum level based on Ly$\alpha$, Ly$\beta$ or Ly$\gamma$ measurements have a direct impact on the models used to describe the reionization process~\citep{fan06}. 

For the case of quasars, the cause of uncertainties can be controlled by improving both the quality and the quantity of the data. The best analysis for measuring the redshift evolution of the Gunn-Peterson optical depth is based on some 19 SDSS quasars at $5.7 < z < 6.4$~\citep{fan06}. As the SDSS survey is not optimized for the detection of high-z quasars (due to its wavelength coverage), currently operating \cite[e.g. UKIDSS,~][]{hewett06} and forthcoming \cite[VISTA,~][]{vista04} infrared surveys shall identify numerous quasar candidates at high redshift. NIRSpec observations will allow both (a) to confirm (or reject) the nature of the proposed candidates and (b) to study the spectral properties of quasars at $z > 7.2$ (for which the Ly$\alpha$ enters NIRSpec's range), to obtain estimates of the fraction of neutral hydrogen as a function of redshift.

\subsection{The AGN unification scheme}

Despite its small aperture (0.8\,m), Spitzer has revolutionised our knowledge about AGN: large-area MIR photometric surveys  revealed numerous dust-obscured AGN candidates~\citep{lacy04, eva05, stern05, ah06, daddi07, donley07}, many of which, due to their very red colours, would have otherwise been missed by optical surveys, due to selection biases. A significant fraction of these objects would even be missed by deep, X-ray surveys, due to the very high column densities of ionized gas surrounding the BH. Using ground-based facilities, optical spectroscopic follow-up observations have been done only for the brightest objects~\citep{lacy07}, as a large fraction of these IR-selected AGN candidates are too faint to obtain spectra with ground-based telescopes.

Selecting AGN using complementary selection criteria is the only way to understand the reason for which the so-called unification scheme~\citep{anton93} does not work for some 30\% of AGN~\cite[][and references therein]{garcet07}. 
According to the unification scheme, the absorption (in X-rays) and extinction (in UV, optical) is a geometrical effect: 
the absorbing material that surrounds the BH (the putative ``torus") might be affecting the observed properties of AGN depending on the viewing angle of the observer with respect to the ``torus". Objects with strong X-ray absorption, however, seem to have mild optical and UV extinction~\citep{page01}, while AGN with narrow emission lines in optical spectra show mild (or even absent) X-ray absorption~\citep{pan02}. Finally, there is observational evidence~\citep{punsly06} that even challenges the fact that broad absorption line (BAL) quasars are X-ray weak~\citep{gal06}.

MIRI shall be an excellent instrument for studying the characteristics of the obscuring material (that mainly emits in the MIR and FIR), which is the key to understanding the controversy between unified scheme predictions and observations. 
The 9.7$\,\mu$m silicate feature, for example, provides an estimate of the absorption due to the ``torus"  and can be associated to the column density N\rm{$_{H}$}, a measure of the obscuration that X-rays experience due to the ionized gas~\citep{li01, shi06}. High-quality MIR spectra will permit to properly measure the continuum level around this feature and associate it to the flux density for unabsorbed continuum, for objects out to $z \approx 1.9$.

\subsection{\label{sb-agn}The starburst - AGN relation}

Sensitive observations with Spitzer are showing that AGN are common in infrared bright galaxies at high-z~\cite[e.g.][]{yan05,dey08} and especially in massive galaxies~\citep{daddi07}.  However, the relationship between AGN and star formation locally and at high redshift is still not well understood. Mergers and strong interactions are believed to trigger AGN activity in galaxies~\citep{heckman86}, and these events can also produce instabilities in the ISM and trigger intense episodes of star formation or starbursts~\citep{ho05}. Nevertheless, it is still not clear how gaseous material is funnelled into the central region of galaxies and what is the relation between star formation and quasar activity. For instance, AGN are located in the transition region (the so-called green valley) between the blue cloud (star-forming galaxies) and the red sequence~\cite[quiescent galaxies, see e.g.][]{nandra07}, and it has been suggested they may play a role in quenching star formation in galaxies. However, so far it is not clear the role of AGN in quenching or triggering star formation~\cite[see, e.g.][]{ah08,bundy08}. 

The IR spectral range is exceptionally rich in features which can be used to differentiate between AGN activity and star formation, such as (a)\,PAH features, (b)\,extended hydrogen recombination line emission, (c)\, the presence of a strong hot dust continuum and (d)\,high-excitation and coronal emission lines~\cite[see, e.g.~][]{sturm02}. All these features can be observed with NIRSpec and MIRI for galaxies at different redshifts, and these observations will allow us to unravel the mechanisms responsible for  star formation and AGN activity in galaxies, and understand their evolution as a function of redshift.

{\it Acknowledgements:} The MIRI European Consortium thanks the following National Funding Agencies for their support of the MIRI: Belgian Science Policy Office, Centre Nationale D'Etudes Spatiales, Danish National Space Centre, Deutsches Zentrum fur Luftund Raumfahrt, Enterprise Ireland, Ministerio De Education y Ciencia, Nova, Science and Technology Facilities Council, Swiss Space Office, Swedish National Space Board. The authors would also like to acknowledge the MIRI science team for the fruitful discussions.



\end{document}